\documentstyle[12pt]{article}

\protect
\protect
\newcommand{\lgm}{{\,\rm ln }}

\renewcommand{\arraystretch}{1.5}

\newcommand{\ice}[1]{\relax}
\newcommand{\ep}{\epsilon}

\newcommand{\bea}{\begin{eqnarray}}
\newcommand{\eea}{\end{eqnarray}}

\newcommand{\prd}{\partial}
\newcommand{\beq}{\begin{equation}}
\newcommand{\eeq}{\end{equation}}

\newcommand{\apis}{\frac{\dsp\alpha_s(s)}{\dsp \pi}}

\newcommand{\ba}{\begin{array}} 
\newcommand{\ea}{\end{array}} 
 
\newcommand{\as}{\alpha_s}
\newcommand{\gm}{\gamma_m}
\newcommand{\G}{\Gamma}
\newcommand{\g}{\gamma}

\newcommand{\gssq}{\gamma^{{\rm  SS}}_q}

\newcommand{\dmu}{\mu^2\frac{d}{d\mu^2}}
\newcommand{\msbar}{\overline{\mbox{MS}}}
\newcommand{\dsp}{\displaystyle}
\newcommand{\EQN}{\label}
\newcommand{\ovl}{\overline}

\begin{document}

\begin{titlepage}

\begin{flushright}
\begin{tabular}{l}
  MPI/PhT/96-61\\
  hep-ph/9608318\\
  August   1996   
\end{tabular}
\end{flushright}
\mbox{}

\protect\vspace*{.3cm}

\begin{center}
  \begin{Large}
  \begin{bf}
Correlator of the quark scalar currents and $\Gamma_{tot}(H \to \mbox{hadrons})$
at ${\cal O}(\as^3)$ in  pQCD  
  \\
  \end{bf}
  \end{Large}
  \vspace{1cm}
K.G.~Chetyrkin $^{a,b}$,
\begin{itemize}
\item[$^a$]
 Institute for Nuclear Research,   
 Russian Academy of Sciences,   \\
 60th October Anniversary Prospect 7a 
 Moscow 117312, Russia 
\item[$^b$]{%
Max-Planck-Institut f\"ur Physik, Werner-Heisenberg-Institut, \\ 
F\"ohringer Ring 6, 80805 Munich, Germany}
\end{itemize}
\vspace{0.2cm}

  \vspace{0.5cm}
  {\bf Abstract}
\end{center}
\begin{quotation}
\noindent
We present the results of the analytical evaluation of the massless
four-loop ${\cal O}(\as^3)$ correction to the correlator of the quark
scalar currents and the Higgs decay rate into hadrons. 
In numerical form 
we found  (in $\ovl{\mbox{MS}}$ scheme)     that
$
\Gamma(H \to  \ovl{b} {b})
= 
\frac{3G_F}{4\sqrt{2}\pi}M_H m_b^2(M_H)
\left[
 1 + 5.667 \alpha_s/\pi
+ \left(35.94 - 1.359 n_f\right)(\alpha_s/\pi)^2
\right.
$
\\
$
\left.
+\left(164.139 - 25.771 n_f + 0.259 n_f^2\right)(\alpha_s/\pi)^3
\right]
$
where  $n_f$ is   the number of  quark favours and 
$\alpha_s = \alpha_s(M_H)$.
\end{quotation}
\vfill

\ice{
In[24]:=  Collect[N[Ex[RSm0as3/3]/.qcd/.Zrule/.{LMS->0},8],as]

                               2
Out[24]= 1. + 5.6666667 as + as  (35.939961 - 1.3586507 nf) + 
 
       3                                          2
>    as  (164.13921 - 25.771192 nf + 0.25897429 nf )
}
\noindent
email:
\\
chet@mppmu.mpg.de

\end{titlepage}

\section{Introduction}
In the Standard Model (SM) one physical scalar  
 Higgs boson is present as a remnant of 
the mechanism of mass generation.
  Particularly interesting for the observation
 of the Higgs boson with  an
  intermediate mass
$M_H<2M_W$ is the dominant decay channel 
into a bottom pair 
$H\rightarrow b\bar{b}$.
( Standard Model
 properties
of the Higgs boson have been discussed 
in many reviews; see for example
\cite{GunHabKanDaw90,Kni93}).
The  partial width
$\Gamma(H\rightarrow b\bar{b})$ is 
significantly affected by QCD
radiative  corrections. 
First order $\alpha_s$ corrections
including the full $m_b$ dependence
were studied by several groups 
\cite{BraLev80,Sak80,InaKub81,DreHik90,KatKim92}. 
Second order corrections were calculated in the 
limit $m_b^2 \ll M_H^2$. Apart from the trivial 
overall factor $m_b^2$ due to the Yukawa coupling,
corrections were obtained for otherwise massless
quarks  
in Refs.~\cite{GorKatLarSur90} and for 
a nonvanishing  mass of the virtual top quark 
in Ref.~\cite{Kni94} (both  results are confirmed in Ref.~\cite{me96}).
Subleading   corrections in the
$m_b^2/M_H^2$ expansion were found in 
Ref.~\cite{Sur94a} and also confirmed in Ref.~\cite{me96}. 
In the latter work  an additional quasi-massless (and numerically
important) contributions of order $\alpha_s^2$ have been
identified and elaborated.  They come from so-called singlet diagrams
with a non-decoupling top quark loop inside (similar effects for the
decay of the Z-boson to quarks were first discovered in \cite{KniKue90b}).
The results of Ref.~\cite{me96} have been confirmed and extended
by  computing extra  terms in the expansion in $M_H^2/m_t^2$
in Ref.~\cite{LarRitVer95}.

In this work we compute the next-next-to-leading massless 
correction of order $\alpha_s^3$ to $\G(H \to \mbox{hadrons})$.

\section{Preliminaries}
We start with considering the two-point correlators
\beq 
\EQN{Pi}
 \Pi^{\rm S}_{ff'}(Q^2)
= (4\pi)^2 i\int dx e^{iqx}\langle 0|\;T[\;J^{\rm S}_f(x)
J^{\rm S}_{f'}(0)\,]\;|0\rangle
{}.
\eeq
Here $Q^2 = - q^2$, $J^{\rm S}_f=\bar{\Psi}_f\Psi_f$ is the scalar current for quarks with
flavour $f$ and mass $m_f$, which are coupled to the scalar Higgs
bosons.  The total hadronic decay rate of a scalar Higgs boson is naturally
expressed in terms of the absorptive part of the correlator (\ref{Pi})
\[
R^{\rm S}_{ff'}(s)
 = \frac{1}{2\pi s}
\mbox{\rm Im} \,  \Pi^{\rm S}_{ff'}(-s-i\epsilon) 
{},
\]
as follows \cite{BraLev80,Sak80,InaKub81}:
\beq \EQN{i3}
\G(H \to \mbox{hadrons})
=\frac{G_F}{4\sqrt{2}\pi}M_H 
\sum_{f f'}m_f m_{f'} R^{\rm S}_{ff'}(s = M_H^2)
\label{decay_rate_from_R}
{}.
\eeq
Assuming that the Higgs boson mass $M_H$ is less then $2m_t$ and much
larger than $m_b$, we shall treat correlators (\ref{Pi}) within the
massless five-quark QCD. (In fact, even in the limit of 
heavy top  the Higgs boson coupling with the current $J^{\rm S}_t$
still contributes to $\G_H$; we shall ignore    these effects in 
our  work. For a recent discussion see e.g. Ref.~\cite{Djuadi96}.) 
 Under such an assumption the nondiagonal correlators
$\Pi^{\rm S}_{ff'}$ with $f \not = f'$ vanish identically while the
diagonal correlators get flavour independent, that is
\[
\Pi^{\rm S}_{ff}(Q^2) \equiv \Pi^{\rm S}(Q^2)
\ \ \mbox{and } \ \ \ 
R^{\rm S}_{ff}(s)\equiv R^{\rm S}(s)
{}.
\]
The renormalization mode of the polarization operator
$\Pi^{\rm S}(Q^2)$  reads  (see, e.g. Ref.~\cite{review})
\beq
\Pi^{\rm S}(Q^2,a_s,\mu^2) = Z^{\rm SS}_q Q^2  + Z_m^2\Pi^{\rm S}_0(Q^2,a_s^0)]
{},
\label{renorm:mod}
\eeq
where $a_s = \alpha_s/\pi= g^2/(4\pi^2)$, $g$ is
the  strong coupling constant;
$Z_{m}$ is the quark mass renormalization constant. 
Within the
$\msbar$ scheme \cite{MSbar}
\beq
Z^{\rm SS}_q = \sum_{1\le j\le i} \left(Z^{\rm SS}_q \right)_{ij} \frac{a_s^{i-1}}{\ep^j}
{}\, 
\label{ZSS}
\eeq
and the coefficients $\left(Z^{\rm SS}_q\right)_{ij}$  are  just numbers,
with  $D=4-2\ep$ standing for the space-time
dimension.
As a result we arrive at  the following renormalization group (RG) equation
for  the polarization operator $\Pi^{\rm S}(Q^2)$
\beq
\label{rgea}
\left(
\mu^2\frac{\partial}{\partial\mu^2}
 +
a_s
\beta(a_s) \frac{\partial}{\partial a_s}
+
2 \g_m(a_s)
\right)
           \Pi^{\rm S} =   \g^{\rm SS}_q(a_s)
{}
\eeq
or, equivalently, ($L_Q = \ln\frac{\mu^2}{Q^2}$)
\beq
\frac{\prd }{\prd L_Q} \Pi^{\rm S}  =
\g^{\rm SS}_q(a_s)
-\left(
  2\g_m(a_s)
+ \beta( a_s) a_s\frac{\prd }{\prd a_s}
\right) \Pi^{\rm S} 
\label{rgPi2}
{}.
\eeq
Here
the  anomalous dimensions $\g^{\rm SS}_q(a_s)$ and $\g_m(a_s)$, 
and the  $\beta$-function $\beta(a_s)$  are  defined in the usual way
\beq
\g^{\rm SS}_q = \mu^2  \frac{d}{d \mu^2}(Z^{\rm SS}_q)-\epsilon
Z^{\rm SS}_q =
-\sum_{i\geq 0}(i+1)
(Z^{\rm SS}_q)_{i1}
a_s ^{i}
{} ,
\label{eq7}
\eeq 
\beq
\dmu a_s  = \as \beta(a_s) \equiv
-\sum_{i\geq0}\beta_i a_s^{i+2}\, ,
\ \ 
\dmu \bar{m}(\mu) =  \bar{m}(\mu)\gm(a_s) \equiv
-\bar{m}\sum_{i\geq0}\gm^i\left( a_s \right)^{i+1}
\label{eq8}
{}.
\eeq
The relation (\ref{rgPi2}) demonstrates  explicitly  the main
computational advantage of finding at first the polarization function
$\Pi^{\rm S}(Q^2)$ against a direct calculation of $R^{\rm S}(s)$ in the case of
massless pQCD.  Indeed, in order $a_s^3$ the derivative $\frac{\prd
}{\prd L_Q} \Pi^{\rm S}$ and, consequently, $R^{\rm S}(s)$ depends on the very
function $\Pi^{\rm S}$ which is multiplied by at least one factor of
$a_s$. Thus, one needs only to know $\Pi^{\rm S}$ to order  $a_s^{2}$ and the
anomalous dimension $\g^{\rm SS}_q(a_s)$ to order  $a_s^{3}$  to
find   all $Q$-dependent terms in $\Pi^{\rm S}$ at ${\cal O}(a_s^3)$,
since  the beta function and the quark mass anomalous dimension
$\gamma_{m} $ are reliably known to order $a_s^3$ from 
Refs.~[17-20].

As we shall see later both problems are eventually reduceable to
calculation of a specific class of diagrams representing some massless
Feynman integrals (FI)  depending on only one external momentum (to be
named {\em p-integrals} below) and with the number of loops not
exceeding three.  Note that this is in an obvious disagreement with a
statement from   Ref.~\cite{Sur94a}
about the inevitable necessity to compute {\em finite} parts of {\em
four-loop} p-integrals in order to find  the ${\cal O}(a_s^3)$ 
contribution to $R^{\rm S}(s)$.  Yet, we agree with the author of
Ref.~\cite{Sur94a} that if such a point were correct it would
certainly preclude, at least at the present state of the art, any
possibility of  analytical calculation of $R^{\rm S}(s)$ to
order $\alpha_s^3$.

\section{Calculation of  $\Pi^{\rm S}$ }
In order $\alpha_s^2$ the polarization operator $\Pi^{\rm S}$ is contributed
by twelve three-loop p-integrals.  Such a  calculation 
is now to be considered as almost trivial one 
because of three facts:
\begin{description}
\item[a]
There is  an elaborated algorithm 
which provides a way  to evaluate  analytically
divergent as well as finite parts of any three-loop dimensionally regulated
 p-integral
\cite{me81a,me81b}. 
\item[b]
The algorithm is   
reliably implemented in the language of FORM 
\cite{Ver91} as the package named MINCER in
Ref.~\cite{mincer2}. 
\item[c] The package has been extensively tested 
and a  chance of  a bug in it seems to be
very  small. 
\end{description}

We have computed the $\Pi^{\rm S}_0$ to $\alpha_s^2$ using the general gauge
with the gluon propagator 
$(g_{\mu\nu} - \xi \frac{q_\mu q_\nu}{q^2})/q^2$. 
On performing the renormalization
according to eq.~(\ref{renorm:mod}) we have obtained the following result:
\begin{eqnarray}
\rule{0.0mm}{0.0mm}\Pi^{\rm SS}(Q^2)&=& 
d[R] Q^2\left\{
-4 
-2  \lgm\frac{\mu^2}{Q^2}
+ a_s
\,C_F 
\left[
-\frac{131}{8} 
+6  \,\zeta(3)
-\frac{17}{2}  \lgm\frac{\mu^2}{Q^2}
-\frac{3}{2}  \lgm^2\frac{\mu^2}{Q^2}
\right]
\right.
\nonumber\\
+a_s^2
\left[
\rule{.0mm}{0.6cm}
\right.
&{}&
\!\!\!\!\!\!
C_F^2
\left(
-\frac{1613}{64} 
+24  \,\zeta(3)
-\frac{9}{4}  \,\zeta(4)
-15  \,\zeta(5)
-\frac{691}{32}  \lgm\frac{\mu^2}{Q^2}
\nonumber
\right.
\\
&{}&
\phantom{MM}
\left.
+\frac{9}{2}  \,\zeta(3) \lgm\frac{\mu^2}{Q^2}
-\frac{105}{16}  \lgm^2\frac{\mu^2}{Q^2}
-\frac{3}{4}  \lgm^3\frac{\mu^2}{Q^2}
\right)
\nonumber\\
\!\!\!\!\!\!
&{+}&\!\!\!\!
\,C_F \,C_A 
\left(
-\frac{14419}{288} 
+\frac{75}{4}  \,\zeta(3)
+\frac{9}{8}  \,\zeta(4)
+\frac{5}{2}  \,\zeta(5)
-\frac{893}{32}  \lgm\frac{\mu^2}{Q^2}
\right.
\label{PiS2:result}
\\
\nonumber
&{}&
\phantom{MMM}
+
\left.
\frac{31}{4}  \,\zeta(3) \lgm\frac{\mu^2}{Q^2}
-\frac{71}{12}  \lgm^2\frac{\mu^2}{Q^2}
-\frac{11}{24}  \lgm^3\frac{\mu^2}{Q^2}
\right)
\\
\!\!\!\!\!\!
&{+}&\!\!\!\!
\,C_F \,T \,n_f 
\left.
\left.
\left(
\frac{511}{36} 
-4  \,\zeta(3)
+\frac{65}{8}  \lgm\frac{\mu^2}{Q^2}
-2  \,\zeta(3) \lgm\frac{\mu^2}{Q^2}
+\frac{11}{6}  \lgm^2\frac{\mu^2}{Q^2}
+\frac{1}{6}  \lgm^3\frac{\mu^2}{Q^2}
\right)
\right]
\right\}
{}.
\nonumber
\end{eqnarray}
Here $a_s = \alpha_s(\mu)/\pi$, $C_A$ and $C_F$  
are the Casimir operators of the adjoint and
quark (defining) representations of the colour group; $T$ is the
normalization of the trace of generators of quark representation
$Tr(t^a t^b) = T \delta^{ab}$; $ n_f $ is the number of quark
flavours; $d[R]$ is the dimension of the quark representation of the
colour group.  Note, that all the Q-dependent terms in
(\ref{PiS2:result}) are in agreement with the results of
\cite{GorKatLarSur90}. The independence of (\ref{PiS2:result}) from
the gauge parameter $\xi$ is of course expected on general grounds.

\section{Calculation of the anomalous dimension $\gssq$ }
There is about a hundred of  diagrams contributing to 
 $\Pi^{\rm S}$ in  order $\alpha_s^3$.
According to (\ref{renorm:mod}), to compute $\gssq$  one should
determine the UV counterterms (that is essentially divergent parts) of
all the corresponding  four-loop p-integrals. Unfortunately, at present
there exists no way to compute  directly the divergent part of a
generic four-loop p-integral.  

The only  available  (unfortunately rather involved) approach to perform such
calculations analytically is to use the method of Infrared
Rearrangement (IRR) discovered by A.~Vladimirov in
Ref.~\cite{Vladimirov80} (see also Refs.~\cite{me79}).  The method
simplifies calculation of UV counterterms by the effective use of the
following important theorem: in a class of minimal renormalization
schemes (including $\msbar$- and MS-ones) any UV counterterm has to be
polynomial in external momenta {\em and} masses
\cite{Collins75}. It amounts to an appropriate transformation 
of the IR structure of FI's by expanding the latter 
in a formal Taylor series with respect to some external momenta and
masses, with resulting FI's being much simpler to calculate.

The method of IRR was significantly extended with elaborating a
so-called $R^*$-operation in Refs.~\cite{me82,me84,me91}.  By an explicit
construction of the corresponding algorithm, it has been shown in
Ref.~\cite{me84} how to reduce calculation of the UV counterterm for an
arbitrary (h+1)-loop FI to evaluation of divergent and finite parts of
some appropriately chosen h-loop p-integrals.

It should be stressed that the $R^*$-operation is absolutely essential
for the algorithm to work in general case, though in most (but not in
all) practical cases one could proceed without it.  However, such a
practice forces the use of a diagram-wise renormalization procedure; 
the latter, being very difficult to perform by a computer, implies a
huge  amount of highly error-prone and time-consuming  
manipulations with  hundreds of diagrams. 

For instance, the only QCD four-loop problem performed by now is the
evaluation of the ratio 
$   R(s) =\sigma_{\mbox{tot}}(e^{+} e^- \to \mbox{hadrons}) /
  \sigma(e^+ e^- \to \mu^+ \mu^-)
$ 
to order $\alpha_s^3$. It was done in
Refs.~\cite{GorKatLar91,SurSam91} within the diagram-wise
renormalization approach.  The core of the problem is the calculation
of the four-loop contribution to the photon anomalous dimension
entering into the RG equation for the photon polarization
operator. The initial 98 four-loop diagrams contributing to  the
photon polarization operator proliferate to about 250 after the IRR
procedure is applied. In addition these diagrams  contain about 600
various subdiagrams which should be computed separately in order to 
subtract UV subdivergences.

Technically, the calculation of $\gssq$ is obviously very similar to
that of  the photon anomalous dimension. Consequently, any attempt
of straightforward repetition of those calculations for the case of
the scalar correlator would mean a few man-years of routine and boring
work and, thus, would  be  not  acceptable for the present author.

Below we  will use, instead, the  power of the $R^*$-operation and
simplify the application of IRR to the problem so far that
both UV and IR renormalizations can be done in a global form and,
consequently, can be simply performed by computer.

We begin with from the Dayson--Schwinger equation for the correlator
(\ref{Pi}) written in the bare form\footnote{For simplicity we 
set  the ${}'$t Hooft-Veltman unit of mass $\mu$ equal to $1$ below.}
\beq
 \Pi^{\rm S}_0 (q,a^0_s)  =-\int \mbox{\rm d} p  \frac{ (4\pi)^2}{(2\pi)^D}
Tr[ G^0(p+q,a^0_s)\G^0_S(p,q,a^0_s) G^0(p,a^0_s)]
{}.
\label{DSeq:bare}
\eeq
Here $G^0$ and $\G^0_S$ are the full quark propagator and the scalar
current vertex function respectively; below the integration with
respect to the loop momentum $ p $ with the weight function 
$\frac{ (4\pi)^2}{(2\pi)^D}$ 
will be only understood but not explicitly displayed.

The renormalized version of (\ref{DSeq:bare}) reads 
\beq
\Pi^{\rm S} (q,a_s)  =
Z^{\rm SS}_q Q^2  
- 
\frac{Z^2_S}{Z_2^2}
Tr[G^0(p+q,a^0_s)\G^0_S(p,q,a^0_s) G^0(p,a^0_s)]
{}.
\label{DSeq:ren1}
\eeq
Here $Z_2$ is the quark wave function renormalization constant;  
$Z_{\rm S}$  is the renormalization constant of the scalar 
quark  current defined as
\[
[\ovl{\psi} \psi ] = Z_{\rm S}/Z_2 \,\,\ovl{\psi^0}\psi^0
{},
\]
where  the current inside squared   brackets is the
renormalized one. A well-known equality
$Z_{\rm S} =  Z_m Z_2 $ implies the equivalence of (\ref{renorm:mod}) and 
(\ref{DSeq:ren1}).

{}From the finiteness of the renormalized correlator one  gets
\begin{eqnarray}
Z^{\rm SS}_q &=& - K_\ep \left\{  
\frac{1}{2 D}\left(\frac{Z_{\rm S}}{Z^2_2} 
\Box_q Tr[ G^0(p+q,a^0_s)\G^0_S(p,q,a^0_s) 
\tilde G^0(p,a^0_s)]
\right.
\right.
\nonumber
\\
&+&
\left.
\left.
\frac{\delta Z_{\rm S} Z_{\rm S}}{Z^2_2} 
\Box_q Tr[ G^0(p+q,a^0_s)\G^0_S (p,q,a^0_s) 
G^0(p,a^0_s)]
\right)
\right\}
{}\, 
\label{Zph1}
\end{eqnarray}
where $K_\ep f(\ep)$ stands for the singular part of the Laurent
expansion of $f(\ep)$ in $\ep$ near $\ep=0$ and 
$\delta Z_{\rm S} = Z_{\rm S} - 1 $.
In Eq.~(\ref{Zph1}) 
we have let a Dalambertian with respect to the external momentum 
$q$ act on quadratically divergent diagrams to transform them to the
logarithmically divergent ones.  We also have introduced an auxiliary
mass dependence to a quark propagator --- the one entering into the
``left''  current $J_S$--- by making the following replacement
\beq
G^0(p,a^0_s) \to \tilde G^0(p,a^0_s) \,\,   p^2/({p^2-m_0^2})
{}.
\label{gamma_tilde}
\eeq
Note, please, that the auxiliary mass dependence has caused somewhat
more complicated structure of UV renormalizations in the right hand
side of Eq.~(\ref{Zph1}).

The idea of the method of IRR is quite simple: since the
renormalization constant $Z^{\rm SS}_q$ does not depend on anything
dimensionful  one could significantly simplify its  calculation  
 by nullifying the momentum $q$ in Eq.~(\ref{Zph1}).
The only requirement which must be respected is the
absence of any IR singularities in the resulting 
integrals. Unfortunately, a mass introduced to a propagator
distinguished by some topological property like the one we created
above is not always sufficient to suppress all IR divergences in all
diagrams.  For instance, if $q=0$ then there appear completely
massless tadpoles in the second term on the rhs of Eq.~(\ref{Zph1}).
Thus, the eq. (\ref{Zph1}) with $q=0$  is {\em not valid}  unless the
unwanted IR poles are all identified and subtracted away.  

This is certainly  the  job  the $R^*$-operation
was  created for! The  rules of Ref.~\cite{me84}  spell how to do it 
on the diagram-wise level. The only remaining problem is to disentangle
the relevant combinatorics and  put down the IR subtractions in a global form.
The task  is facilitated by the fact that, as shown in
Ref~.\cite{me91} the IR counterterms for an arbitrary  diagram can 
be determined in terms of some properly  chosen combination of the UV ones.
The final formula incorporating all necessary  UV and IR subtractions
reads 
\begin{eqnarray}
Z^{\rm SS}_q &=& - K_\ep \left\{  
\frac{1}{2 D} 
\frac{Z_{\rm S}}{Z^2_2}
\Box_q Tr[ G^0(p+q,a^0_s)\G^0_S (p,q,a^0_s) 
\tilde G^0(p,a^0_s)]|_{{}_{\displaystyle q=0}}
\right.
\nonumber
\\
&-& 
\left.
\frac{Z_{\rm S}}{Z^2_2}
\frac{1}{4}Tr[\delta\G^0_{\tilde S}(0,0,a^0_s) ] 
\frac{Z^{\rm SS}_q}{Z^2_m}
-
\frac{\delta Z_{\rm S } Z_{\rm S}Z^{\rm SS}_q }{Z^2_2 Z^2_m}
\right\}
{}\, .
\label{final_eq}
\end{eqnarray}
Here,  by $\delta\G^0_S(p,q,a^0_s)$
we denote the vertex function of the scalar  current
with the tree contribution removed. The ``tilde'' atop  $S$ 
again  means that  in every diagram the quark propagator 
entering to the vertex $J^{\rm S}$ is softened at small momenta by
means of the auxiliary mass $m_0$  according to Eq.~(\ref{gamma_tilde}).
The bare coupling constant  $a_s^0$ is to be understood as 
$a_s = Z_a a_s$,  with  $Z_a$ being 
the coupling constant renormalization constant.  
To our accuracy in $a_s$ 
\[
Z_a = 1 + a_s(-\beta_0/\ep) + a_s^2(\beta_0^2/\ep^2 -\beta_1/(2\ep))
{}\,.
\]
Finally,  an inspection of (\ref{final_eq}) immediately shows that,
in order to find the $(n+1)$-loop  correction to $Z^{\rm SS}_q$, 
one  needs only to  know the renormalization constants $Z^{\rm SS}_q$, 
$Z_2$ and $Z_m$  to order  $a_s^n$ as well as    the bare Green functions 
\beq
 G^0(p,a^0_s), \ \ \frac{\partial}{\partial q_\beta }
[\G^0_S (p,q,a^0_s)]|_{{}_{\displaystyle q=0}} 
, \ \ 
\Box_q [\G^0_S(p,q,a^0_s)]|_{{}_{\displaystyle q=0}}, 
\ \ 
\delta\G^0_{\tilde S}(0,0,a^0_s)
\label{functions}
{}
\eeq
up to (and including) $n$-loops.  Thus,  we have
obtained a general formula for $Z^{\rm SS}_q$ in terms of bare
\mbox{$p$-integrals} with explicitly resolved UV and IR subtractions.

\section{Results and discussion}
We have computed with the program MINCER \cite{mincer2} 
the unrenormalized three-loop
Green functions (\ref{functions}) as well as the  
renormalization constants $Z_m$ and $Z_2$ to order $a_s^3$.  The
calculations have been performed in the general covariant gauge.  The
total calculational time with an IBM  workstation is about 40 hours
for the general gauge; for the Feynman one it is reduced to about 4
hours.

Then we have used Eqs.~(\ref{final_eq}) and (\ref{eq7})  
to find $\g^{\rm SS}_q$  to order $a_s^3$. 
Our results for $\g^{\rm SS}_q$   and  $R^{\rm S}(s)$  read
\begin{eqnarray}
\gamma^{\rm SS}_q &{}& =  \nonumber
\\
\!\!\! d[R]&{}& \!\!\!\!\!\!\!\!\!\!\!\! \left\{
\rule{.0mm}{0.6cm}
\right.
\nonumber
-2
+a_s  \,C_F 
\left[
 -\frac{5}{2}
\right]
+
a_s^2
\left[
 C_F^2
\left(
\frac{119}{32} 
-\frac{9}{2}  \,\zeta(3)
\right)
{+} \,C_F \,C_A 
\left(
-\frac{77}{16} 
+\frac{9}{4}  \,\zeta(3)
\right)
{+} \,C_F \,T \,n_f 
\right]
\nonumber\\
\nonumber
+a_s^3
\left[
\rule{.0mm}{0.6cm}
\right.
\!\!\!\!
&{}& C_F^3
\left(
-\frac{4651}{384} 
-\frac{29}{4}  \,\zeta(3)
+\frac{27}{8}  \,\zeta(4)
+\frac{45}{4}  \,\zeta(5)
\right)
\\
\nonumber
&{+}& C_F^2\,C_A 
\left(
\frac{641}{48} 
-\frac{259}{16}  \,\zeta(3)
+\frac{39}{16}  \,\zeta(4)
+\frac{45}{8}  \,\zeta(5)
\right)
\nonumber\\
&{+}& \,C_F C_A^2
\left(
-\frac{267889}{31104} 
+\frac{475}{48}  \,\zeta(3)
-\frac{33}{16}  \,\zeta(4)
-\frac{45}{8}  \,\zeta(5)
\right)
\nonumber
\\
\nonumber
&{+}& 
C_F^2\,T \,n_f 
\left(
\frac{125}{32} 
-\frac{1}{2}  \,\zeta(3)
-3  \,\zeta(4)
\right)
\nonumber\\
&{+}& \,C_F \,C_A \, T \,n_f 
\left(
\frac{631}{7776} 
+\frac{5}{3}  \,\zeta(3)
+\frac{9}{4}  \,\zeta(4)
\right)
\nonumber
\\
&{+}&
\,C_F  T^2 \, n_f^2
\left(\frac{1625}{1944} 
-\frac{2}{3}  \,\zeta(3)
\right)
\left.
\left.
\rule{.0mm}{0.6cm}
\right]
\rule{.0mm}{0.6cm}
\right\}.
\label{gssq }
\end{eqnarray}
and
\begin{equation}
\begin{array}{l}
R^{\rm S}(s,\mu) = d[R]\left\{1
+ 
a_s(\mu)
\left[
s_1
+
2\gamma_m^0 \lgm\frac{\mu^2}{s} 
\right]
\right.
\\
\left.
\rule{3.2cm}{0cm}
+
a^2_s(\mu)
\left[
s_2 
+
\lgm\frac{\mu^2}{s} \left(
s_1\beta_0 + 2 s_1 \gamma_m^0 + 2 \gamma_m^1
   \right)
+
\lgm^2\frac{\mu^2}{s} \left(
\beta_0 \gamma_m^0 + 2 (\gamma_m^0)^2
   \right)
\right]
\right.
\\ 
\rule{3.4cm}{0cm}
+
a^3_s(\mu)
\left[s_3
+
\lgm\frac{\mu^2}{s} \left( 2 s_2\beta_0
+s_1 \beta_1 + 2 s_2\gamma_m^0 + 2s_1 \gamma_m^1 + 2 \gamma^2_m
                   \right)
\right.
\\
\left.
\rule{4.6cm}{0cm}
+
\lgm^2\frac{\mu^2}{s} \left(
 s_1\beta_0^2 + 3 s_1\beta_0 \gamma_m^0 + \beta_1\gamma_m^0
+ 2 s_1 (\gamma_m^0)^2 + 2\beta_0 \gamma_m^1 + 4 \gamma_m^0 \gamma_m^1
   \right)
\right.
\\
\left.
\rule{4.6cm}{0cm}
+
\lgm^3\frac{\mu^2}{s} \left(
\frac{2}{3}\beta_0^2 \gamma_m^0 + 2 \beta_0 (\gamma_m^0)^2 
+\frac{4}{3}(\gamma_m^0)^3 
   \right)
\left.
\rule{0.0cm}{0.6cm}
\right]
\right\}
{}.
\end{array}
{}.
\label{RS0}
\end{equation}
Here the coefficients of the beta-function $\beta_0, \ \beta_1$ and
the quark mass anomalous dimension $\gamma_m^i, \ \ i =0,1,2$ are
defined according to (\ref{eq7},\ref{eq8}) and read
\begin{equation} \label{a3}
\begin{array}{lll} \displaystyle
\gamma_m^0
& = & \displaystyle
\frac{1}{4}[3 C_F],
\ \ \ \ \ \  
\gamma_m^1
 =  \displaystyle
\frac{1}{16}\left[
\frac{3}{2} C_F^2 + \frac{97}{6} C_F C_A
- \frac{10}{3} C_F T n_f \right]
{},
\\
{}
\\ 
\displaystyle
\gamma_m^2
& = & \displaystyle
\frac{1}{64}\left[
\frac{129}{2} C_F^3 - \frac{129}{4} C_F^2 C_A
+ \frac{11413}{108} C_F C_A^2 \right.
\\ \displaystyle
{}
\\
&  & \displaystyle
\left.
+C_F^2 T n_f (48 \zeta(3) -46)
+ C_F C_A T n_f
   \left(-48 \zeta(3) -\frac{556}{27}\right)
- \frac{140}{27} C_F T^2 n_f^2 \right]
{},
\end{array}
\end{equation}
\begin{equation} \label{a6}
\beta_0
 =  \displaystyle
\frac{1}{4}\left[
\frac{11}{3} C_A - \frac{4}{3} T n_f \right]
{},
\ \ \ \
\beta_1
 =  \displaystyle
\frac{1}{16}\left[
\frac{34}{3} C_A^2 - 4 C_F T n_f
- \frac{20}{3} C_A T n_f \right]
{}.
\end{equation}
At last,  the coefficients $s_1, \ s_2$ and $s_3$ are found to be
\bea
\dsp
s_1 &=& 
\,C_F 
\left[
\frac{17}{4}\right], 
 \ \ \ 
\nonumber
s_2 = 
C_F^2
\left[
\frac{691}{64} 
-\frac{3}{8}  \pi^2
-\frac{9}{4}  \,\zeta(3)
\right]
{+} \,C_F \,C_A 
\left[
\frac{893}{64} 
-\frac{11}{48}  \pi^2
-\frac{31}{8}  \,\zeta(3)
\right]
\\
\nonumber
&{}&
{+} \,C_F \,T \,n_f 
\left[
-\frac{65}{16} 
+\frac{1}{12}  \pi^2
+
   \,\zeta(3)
\right]
{},
\\
{s_3} &=& \nonumber\\
&{+}& C_F^3
\left[
\frac{23443}{768}
-\frac{27}{16}  \pi^2
-\frac{239}{16}  \,\zeta(3)
+\frac{45}{8}  \,\zeta(5)
\right]
\nonumber\\
&{+}& C_F^2\,C_A
\left[
\frac{13153}{192}
-\frac{383}{96}  \pi^2
-\frac{1089}{32}  \,\zeta(3)
+\frac{145}{16}  \,\zeta(5)
\right]
\nonumber\\
&{+}& \,C_F C_A^2
\left[
\frac{3894493}{62208}
-\frac{1715}{864}  \pi^2
-\frac{2329}{96}  \,\zeta(3)
+\frac{25}{48}  \,\zeta(5)
\right]
\nonumber\\
&{+}& C_F^2\,T \,n_f
\left[
-\frac{88}{3}
+\frac{65}{48}  \pi^2
+\frac{65}{4}  \,\zeta(3)
+\frac{3}{4}  \,\zeta(4)
-5  \,\zeta(5)
\right]
\nonumber\\
&{+}& \,C_F \,C_A \, T \,n_f
\left[
-\frac{33475}{972}
+\frac{571}{432}  \pi^2
+\frac{22}{3}  \,\zeta(3)
-\frac{3}{4}  \,\zeta(4)
+\frac{5}{6}  \,\zeta(5)
\right]
\nonumber\\
&{+}& \,C_F  T^2 \, n_f^2
\left[
\frac{15511}{3888}
-\frac{11}{54}  \pi^2
-  \,\zeta(3)
\right]
{}.
\label{RS}
\end{eqnarray}
We observe that neither $\g^{\rm SS}_{q}$ nor $R^{\rm S}(s)$ depend on the gauge
fixing parameter $\xi$ as it must be.  For the standard QCD  
colour group values $d[R] =3$, $C_F= 4/3, \, C_A = 3$ and $T = 1/2$ we get
for $R^{\rm S}(s)$ with $\mu^2$ set to $s$:
\begin{eqnarray}
R^{\rm S}(s) &=& 3 \left\{\rule{0cm}{0.5cm}
\right.
1
{+} a_s
\left[
 \frac{17}{3}\right]
{+} a_s^2
\left[
\frac{10801}{144}
-\frac{19}{12}  \pi^2
-\frac{39}{2}  \,\zeta(3)
-\frac{65}{24}  \,n_f
+\frac{1}{18}  \pi^2 \,n_f
+\frac{2}{3}  \,\zeta(3) \,n_f
\right]
\nonumber\\
{+} a_s^3
\left[\rule{0cm}{0.5cm}
\right.
&{}&
\frac{6163613}{5184}
-\frac{3535}{72}  \pi^2
-\frac{109735}{216}  \,\zeta(3)
+\frac{815}{12}  \,\zeta(5)
-\frac{46147}{486}  \,n_f
+\frac{277}{72}  \pi^2 \,n_f
\nonumber
\\ &{+}&
\left.
\frac{262}{9}  \,\zeta(3) \,n_f
-\frac{5}{6}  \,\zeta(4) \,n_f
-\frac{25}{9}  \,\zeta(5) \,n_f
+\frac{15511}{11664}  \, n_f^2
\left.
-\frac{11}{162}  \pi^2 \, n_f^2
-\frac{1}{3}  \,\zeta(3) \, n_f^2
\right]
\right\}
{},
\nonumber\\
\EQN{RS3}
\end{eqnarray}
or, in the numerical form, 
\begin{eqnarray}
R^{\rm S}(s) = 3\left\{
\rule{1cm}{0cm}
\right.
 1 &+& 5.66667 \apis
+ \left(35.93996 - 1.35865 n_f\right)\left(\apis\right)^2
\label{RS:numerical}
\\
&{+}&
\left.
\left(164.13921 - 25.77119 n_f + 0.258974 n_f^2\right)\left(\apis\right)^3
\right\}
{}. 
\nonumber
\end{eqnarray}
\ice{
In[22]:=  Collect[N[Ex[RSm0as3/3]/.qcd/.Zrule/.{LMS->0},8],as]

                              2
Out[22]= 1. + 5.666667 as + as  (35.93996 - 1.358651 nf) + 
 
       3                                       2
>    as  (164.1392 - 25.77119 nf + 0.2589743 nf )
     }
At last, for the phenomenologically  relevant case of $n_f = 5 $  we obtain
\beq
R^{\rm S}(s) = 3\left\{
1 
+ 
5.66667 \apis
+ 
29.1467\left(\apis\right)^2
+
 41.7576\left(\apis\right)^3
\right\}
{}. 
\label{}
\eeq
Due to eq.~(\ref{decay_rate_from_R}) the combination $m_b^2 R^{\rm S}(s = M^2_H)$
is directly related to the Higgs decay rate to the   
$ b \ovl{b}  $ pairs plus gluons.  
The corresponding expression including power suppressed corrections
reads
\bea
\Gamma(H \to  \ovl{b} {b})
\nonumber
&=&
\\
\frac{3G_F}{4\sqrt{2}\pi}M_H m_b^2(M_H)
\left[
\rule{1cm}{0cm}
\right.
1
&+&
5.67 \,  a_s(M_H)
+ 
29.15 \,  a_s^2(M_H)
+
41.76   \, a_s^3(M_H)
\nonumber
\\
&{+}& 
\left.
\frac{ m_b^2(M_H)}{M_H^2}
\left(
-6 - 40 \, a_s(M_H) - 87.72 \, a_s^2(M_H)
\right)
\right]
\label{decay_rate_5}
{}.
\eea
Let us take  as input
parameters a $\Lambda_{QCD}^{(5)}=233$ MeV  and a
bottom pole mass of $m^{\rm pole}_b=4.7$ GeV.
The latter translates into the running mass
${m}_b(M_H^2)=2.84/2.75/2.69$ GeV for Higgs
masses of $M_H=70/100/130$ GeV.
All other light quarks are assumed to be massless. 
One arrives at the following
values for the strong coupling constant:
$\as(M_H^2)=0.125/0.118/0.114$ corresponding
to the three different values of $M_H$.
The corresponding numerical values for different
contributions to $R^{\rm S}(M_H^2)$ are given\footnote{
It should be noted that this discussion is somewhat 
inconsistent  because of the following reasons. 
First, the relation between the running and the 
pole  masses is  known only with the $\alpha_s^2$ accuracy
\cite{Bro90}. 
Second, one needs  not yet available coefficients, $\gamma_m^3$
and $\beta_3$,
 for 
taking  into account the  terms of order
$(\alpha_s(m_b^{pole})  -\alpha_s(M_H))^3$
in the running of the quark mass  and the coupling constant. 
Nevertheless,  
the numbers in Table 1 are correct for the given values
of $\alpha_s(M_H)$ and ${m}_b(M_H)$. } 
in Table 1. 
\begin{table}[ht]
\renewcommand{\arraystretch}{1.3}
\begin{center}
\begin{tabular}{|c|c|c|c|c|c|}
\hline
  $M_H$
  & $\alpha_s(M_H)$
  & ${m}_b(M_H^2)$
  & ${\cal O}\left(\alpha_s  (M_H)\right)$
  & ${\cal O}\left(\alpha^2_s(M_H)\right)$
  & ${\cal O}\left(\alpha^3_s(M_H)\right)$
 \\
\hline
\hline
70  GeV& 0.125 & 2.84 GeV & 0.226 - 0.00262  & 0.0461 - 0.00023    & 0.0026     \\
100 GeV& 0.118 & 2.75 GeV & 0.213 - 0.00114  & 0.0411 - 0.00009    & 0.0022     \\
130 GeV& 0.114 & 2.69 GeV & 0.206 - 0.00062  & 0.0384 - 0.00005    & 0.0020    \\
\hline
\end{tabular}
\end{center}
\caption{\label{allmoments}
The contributions of different orders to $R^{\rm S}/3$; the
first number stands for the  massless part  and the second one  for 
the power suppressed ${\cal O}(\frac{m_b^2}{M_H^2})$  part  (where
available). 
         }
\end{table}

\noindent 
To summarize: we have suggested a new convenient way to compute the UV
renormalization constant of the correlator of the scalar quark
currents. Our final formula (\ref{final_eq}) directly expresses the
constant in terms of unrenormalized p-integrals, with all UV and IR
subtractions being implemented in a global form.  The formula is
ideally suited for carrying out completely automatic calculations and
can be easily extended for the case of general bilinear quark
currents.  
Detailed derivation of the formula will be presented elsewhere.
Using the formula and the FORM version of the MINCER we
have computed the ${\cal O}(\alpha_s^3)$ correction to the anomalous
dimension $\gssq$, to the absorptive part of the scalar correlator
$R^{\rm S}(s)$ and to $\G(H \to \mbox{hadrons})$ in pQCD. All the
calculations have been performed with the use of  the general
covariant gauge.  The gauge independence of  the results
constitutes a strong check of the correctness of our approach.

Numerically, the correction of order $\alpha_s^3$ to 
$\G(H\to \mbox{hadrons})$ proves to be relatively small. However, for the 
Higgs boson with intermediate mass it is more important than the 
power suppressed ${\cal O}(\frac{m_b^2}{M_H^2} \alpha^2_s)$
contribution.  (see Table 1).

\vskip0.3cm  

\noindent
{\Large{\bf Acknowledgments}}
\vskip0.3cm

I would like to thank J.~K\"uhn, B.~Kniehl, S.~Larin and
M.~Steinhauser for stimulating and useful discussions. I am specially
grateful to S.~Larin and J.~Vermasseren for providing me with the FORM
package MINCER as well useful advice about its features.  I would like
to thank S. Jadach for the occasion to present the results of this
paper at the Cracow International Symposium on Radiative Corrections,
1-5 August 1996.

I am deeply thankful to the Institute of Theoretical Particle Physics
of the Karlsruhe University and the theoretical group of the
Max-Plank-Institute of  Physics and Astrophysics for the warm
hospitality. Financial
support by Deutsche Forschungsgemeinschaft 
(grants No. Ku 502/3-1 and Ku 502/6-1 ) and by 
INTAS under contract No. INTAS -93-0744 is gratefully acknowledged.

\sloppy\raggedright
\def\app#1#2#3{{\it Act. Phys. Pol. }{   B #1} (#2) #3}
\def\apa#1#2#3{{\it Act. Phys. Austr.}{   #1} (#2) #3}
\def\lhc{Proc. LHC Workshop, CERN 90-10}
\def\npb#1#2#3{{\it Nucl. Phys. }{   B #1} (#2) #3}
\def\plb#1#2#3{{\it Phys. Lett. }{   B #1} (#2) #3}
\def\prd#1#2#3{{\it Phys. Rev. }{   D #1} (#2) #3}
\def\pR#1#2#3{{\it Phys. Rev. }{   #1} (#2) #3}
\def\prl#1#2#3{{\it Phys. Rev. Lett. }{   #1} (#2) #3}
\def\prc#1#2#3{{\it Phys. Reports }{   #1} (#2) #3}
\def\cpc#1#2#3{{\it Comp. Phys. Commun. }{   #1} (#2) #3}
\def\nim#1#2#3{{\it Nucl. Inst. Meth. }{   #1} (#2) #3}
\def\pr#1#2#3{{\it Phys. Reports }{   #1} (#2) #3}
\def\sovnp#1#2#3{{\it Sov. J. Nucl. Phys. }{   #1} (#2) #3}
\def\jl#1#2#3{{\it JETP Lett. }{   #1} (#2) #3}
\def\jet#1#2#3{{\it JETP Lett. }{   #1} (#2) #3}
\def\zpc#1#2#3{{\it Z. Phys. }{   C #1} (#2) #3}
\def\ptp#1#2#3{{\it Prog.~Theor.~Phys.~}{   #1} (#2) #3}
\def\nca#1#2#3{{\it Nouvo~Cim.~}{   #1A} (#2) #3}

\end{document}